%% file: isolating_linear_RSD.tex
\documentclass[fleqn,usenatbib]{mnras}

\usepackage[T1]{fontenc}

\DeclareRobustCommand{\VAN}[3]{#2}
\let\VANthebibliography\thebibliography
\def\thebibliography{\DeclareRobustCommand{\VAN}[3]{##3}\VANthebibliography}

\usepackage{graphicx}	
\usepackage{amsmath}	
\usepackage{amssymb}	

\usepackage{caption}
\usepackage{subcaption}
\usepackage{xcolor}

\usepackage{newtxtext,newtxmath}

\newcommand{\mhmpc}{\,h^{-1}{\rm Mpc}}

\def\xipp{\xi^s(r_{\perp},r_{\parallel})}

\def\({\left(}
\def\){\right)}
\def\[{\left[}
\def\]{\right]}
\def\mhmpc{\,h^{-1}{\rm Mpc}}

\title[Isolating linear RSD signal]{Isolating the linear signal when making redshift space distortion measurements}

\author[M. J. Chapman et al.]{\parbox{\textwidth}{
Michael J. Chapman,$^{1,2}$\thanks{E-mail: mj3chapm@uwaterloo.ca}
Zhongxu Zhai,$^{3,4,1,2}$
Will J. Percival,$^{1,2,5}$
 } \vspace*{4pt}\\
$^{1}$ Waterloo Centre for Astrophysics, University of Waterloo, Waterloo, ON N2L 3G1, Canada \\ 
$^{2}$ Department of Physics and Astronomy, University of Waterloo, Waterloo, ON N2L 3G1, Canada \\
$^{3}$ Department of Astronomy, School of Physics and Astronomy, Shanghai Jiao Tong University, Shanghai 200240, China \\
$^{4}$ Shanghai Key Laboratory for Particle Physics and Cosmology, Shanghai 200240, China \\
$^{5}$ Perimeter Institute for Theoretical Physics, 31 Caroline St. North, Waterloo, ON N2L 2Y5, Canada \\
}

\date{Accepted XXX. Received YYY; in original form ZZZ}

\pubyear{2022}

\begin{document}
\label{firstpage}
\pagerange{\pageref{firstpage}--\pageref{lastpage}}
\maketitle

\begin{abstract}
Constraints on the linear growth rate, $f\sigma_8$, using small scale redshift space distortion measurements have a significant statistical advantage over those made on large scales. However, these measurements need to carefully disentangle the linear and non-linear information when interpreting redshift space distortions in terms of $f\sigma_8$. It is particularly important to do this given that some previous measurements found a significant deviation from the expectation based on the $\Lambda$CDM model constrained by Planck CMB data. We construct a new emulator-based model for small scale galaxy clustering with scaling parameters for both the linear and non-linear velocities of galaxies, allowing us to isolate the linear growth rate. We train the emulator using simulations from the AbacusCosmos suite, and apply it to data from the extended Baryon Oscillation Spectroscopic Survey (eBOSS) luminous red galaxy sample. We obtain a value of $f\sigma_8(z=0.737)=0.368\pm0.041$, in 2.3-$\sigma$ tension with the Planck 2018 $\Lambda$CDM expectation, and find less dependence on the minimum measurement scale than previous analyses.
\end{abstract}




\section{Introduction}

\input{1-introduction}

\section{Modelling RSD including velocity scaling}
\label{sec:isolating}

\input{2-model}

\section{Measuring the eBOSS LRG RSD} 
\label{sec:results}

\input{3-results}

\section{Discussion and Conclusion}
\label{sec:discussion}

\input{4-discussion}

\section*{Acknowledgements}

WP acknowledges the support of the Natural Sciences and Engineering Research Council of Canada (NSERC), [funding reference number RGPIN-2019-03908] MC and WP acknowledge funding from the Canadian Space Agency.

Research at Perimeter Institute is supported in part by the Government of Canada through the Department of Innovation, Science and Economic Development Canada and by the Province of Ontario through the Ministry of Colleges and Universities.

This research was enabled in part by support provided by Compute Ontario (computeontario.ca) and the Digital Research Alliance of Canada (alliancecan.ca).

\section*{Data Availability}

The eBOSS galaxy and random catalogues are publicly available at: https://data.sdss.org/sas/dr16/eboss/lss/catalogs/DR16/ with a description here: https://www.sdss.org/dr16/spectro/lss/ We used the Cobaya package \citep{Torrado:2020xyz, Lewis:2002ah, Lewis:2013hha, Neal:2005}, which is available here: https://github.com/CobayaSampler


\bibliographystyle{mnras}
\bibliography{isolating_linear_RSD.bib} 



\appendix
\section{Pairwise velocities}
\label{app:pairwise-vel}
\input{appendix.tex}


\bsp	
\label{lastpage}
\end{document}

%% file: 1-introduction.tex
Precise measurement of the Cosmic Microwave Background (CMB) has fundamentally changed the way we understand our Universe. We now have tight constraints on the core cosmological parameters and find good agreement with a cosmological model with a cold dark matter component that dominates the matter density and a cosmological constant that dominates the energy density ($\Lambda$CDM). However, there exist several tensions between measurements of the early universe through the CMB and some late time probes. In particular the expansion rate of the Universe at present day, $H_0$, measured in the local Universe from type Ia supernovae \citep{riess22} and the amplitude of fluctuations in matter density field, parameterized as $S_8$, measured by weak lensing surveys \citep{kids1000, desy3} are in disagreement with the values measured from the CMB by the Planck satellite \citep{Planck-2018-overview, Planck-2018-params}. The focus of many ongoing cosmological observations is to build on the current concordance cosmology using additional measurements that are independent of the CMB observations or have complementary parameter degeneracies. To that end, redshift space distortion (RSD) measurements provide a unique test of cosmological constraints derived from the matter density field by probing the velocity field.

RSD is an apparent effect observed in spectroscopic galaxy clustering surveys caused by the peculiar velocities of galaxies. In a spectroscopic galaxy survey the radial distances to the galaxies are usually determined from the redshifts, assuming that the recession velocities are caused entirely by the expansion of the Universe. However, because galaxies have an additional peculiar velocity caused by structure growth and primarily sourced from gravity, their radial positions as determined by the survey, called redshift space positions, will be offset from their true positions in real space \citep{Kaiser:1987}. In the linear regime the amplitude of the velocity field is directly proportional to two cosmological parameters. The first is the logarithmic growth rate of density perturbations, $f$. The second is the amplitude of density fluctuations, which can be normalized using the standard deviation of density fluctuations in a sphere of $8 \mhmpc$, defined as $\sigma_8$. Due to the degeneracy between these parameters RSD constraints are given in terms of the parameter combination $f\sigma_8$ \citep{guzzo_test_2008, song_reconstructing_2009}. RSD measurements can therefore be used to constrain $f\sigma_8$ in a way that is complementary to probes of the density distribution \citep{Huterer:2018}.

RSD measurements are most easily interpreted on linear scales where the density field can be easily modelled analytically (see e.g. \citealt{LRG_corr}). Models can be extended to quasi-linear scales using Lagrangian perturbation theory (LPT), which models the evolution of the density field by the displacement of dark matter fluid elements \citep{taruya2010, reid2011, carlson2013, wang2014}. The perturbation theory expansion breaks down at the shell-crossing scale. An alternative method is the effective field theory (EFT) approach, which makes use of the relatively weak link between the small scale non-linear structure of galaxy formation and the typical separation of galaxies in large scale structure surveys \citep{baumann2012, hertzberg2012}. By integrating out short-wavelength perturbations it becomes possible to solve the resulting smoothed field with a high degree of accuracy into the quasi-linear regime by extending the perturbation theory calculations to arbitrarily high-order \citep{D'Amico:2020, Ivanov:2020, Chen:2021}. While these methods are successful at modelling the distribution of matter in the linear and quasi-linear regimes, they can not provide an analytic basis for the formation of galaxies or the non-linear motion of virialized structures. These effects are instead included as additional correction terms whose functional form can be predicted from perturbation, but with unknown amplitudes that must either be calibrated on simulations or fit from the data \citep{Cabass:2022}.

Previous works have attempted to extract RSD information from small scales by modelling the formation of non-linear structure with N-body simulations. \citet{Reid:2014} used an N-body simulation at a single fixed cosmology to model the clustering of galaxies within the Baryon Oscillation Spectroscopic Survey (BOSS) CMASS sample between $0.8-32\mhmpc$, and found a factor of 2.5 improvement in precision over the perturbation theory RSD analysis on large scales of the same sample. This method has been expanded through the use of machine learning emulators to allow for varying cosmology without needing to run additional N-body simulations for each new point in parameter space, finding similar improvements in precision over perturbation theory approaches \citep{chapman22, Zhai_2021, yuan22, kobayashi22}. 

A key aspect of the \citet{Reid:2014} analysis was the introduction of a velocity scaling parameter, $\gamma_f$, that multiplied all halo velocities in the simulation. Scaling the amplitude of the velocity field is directly equivalent to a proportional change in $f\sigma_8$ in linear theory, allowing \citet{Reid:2014} to specifically assess deviations in the growth rate within a $\Lambda{\rm CDM}$ framework, since the growth rate is normally fixed by the other cosmological parameters. \citet{chapman22} analyzed the eBOSS LRG sample using a Gaussian process based emulator with a velocity scaling parameter \citep{zhai_aemulus3}, however they were forced to restrict their analysis with a minimum scale cut to match the scale where changing $\gamma_f$ no longer directly matched the expectation for a change in $f\sigma_8$. While the small-scale, non-linear velocities are certainly affected by a change in the growth rate, it is no longer necessary that that change be directly proportional, so there is a potential for a systematic bias in applying a linear velocity scaling to non-linear velocities.

This highlights a larger issue in the area of small-scale RSD measurements; how to measure a linear quantity in the non-linear regime without allowing the non-linear velocity evolution to bias the results. This is the primary motivation for this work. We build on the previous model by splitting the velocity scaling parameter $\gamma_f$ into two parameters: $\gamma_l$ to scale the linear component of the velocity, and $\gamma_n$ to scale the non-linear component. This new parameterization allows us to interpret a change in $\gamma_l$ as a change in the amplitude of the linear velocity field consistent with a change in $f\sigma_8$ within a $\Lambda {\rm CDM}$ framework, while $\gamma_n$ allows enough freedom for the non-linear velocity to vary without directly matching the scaling of the linear velocity.

This paper is structured as follows. In Sec.~\ref{sec:isolating} we expand on the model of \citet{chapman22} to isolate the linear signal in the non-linear regime using our new velocity scaling parameters. Then we refit the eBOSS LRG data using the new emulator, and present the results in Sec.~\ref{sec:results}. Finally, in Sec.~\ref{sec:discussion} we discuss the significance of our new results and compare to the work of the previous emulator and other related measurements.

%% file: 2-model.tex
\subsection{Building an emulator}
\label{sec:emulator}

In order to access RSD information on small scales we need to model the clustering of galaxies into the non-linear regime. The solution we choose is to construct an emulator for the small scale clustering, trained and validated using N-body simulations. We apply machine learning with a Gaussian process to emulate the correlation function measurements in each separation bin, as a function of the set of parameters specifying the cosmology and HOD model. First, we use a set of {\it training} data to specify the value of the emulator at a series of points in parameter space. These are the means of the Gaussian distributions. We then use a different set of {\it test} data to optimise the width and shape of an "interpolation kernel", such that the final model given a set of model parameters is the linear sum of the means coming from the training data, weighted by this kernel. Details of the kernel and optimisation are available in \citet{zhai_aemulus3}.
Our training data is generated from N-body simulations, where we use a halo occupation distribution to connect galaxies to halos. While the training data can only have a limited number of possible values in our parameter space, the trained emulator is very effective at interpolating within this parameter space to produce accurate clustering measurements.

In this work we build on the emulator used in \citet{chapman22}, originally based on \citet{zhai_aemulus3}. The emulator used a 5-parameter cosmological model consisting of $\Omega_M$, $\Omega_b$, $\sigma_8$, $h$, and $n_s$, as well as an 8-parameter HOD model to connect galaxies to halos in the simulation, described by the parameters $f_{\rm max}$, $\sigma_{\log M}$, $\log M_{\mathrm{sat}}$, $\alpha$, $\log M_{\mathrm{cut}}$, $c_{\rm vir}$, $v_{\rm bc}$, and $v_{\rm bs}$.

The final parameter of the \citet{chapman22} emulator was a velocity scaling parameter, $\gamma_f$. Physically, $\gamma_f$ rescaled all halo bulk velocities in the simulation, where we define 'bulk velocities' to mean the velocity of the halo as a single unit, rather than the velocity of the individual particles making up the halo or the internal velocity dispersion of the halo. In the linear regime the amplitude of the velocity field is directly proportional to $f\sigma_8$, so a scaling of the velocity field has the same effect as scaling the logarithmic growth rate $f$ \citep{Reid:2014}. However, an issue highlighted in \citet{chapman22} is the question of what velocities can be considered as linear for the purposes of the growth rate. While a change in the growth rate will affect all components of the velocity, the relation between the amplitude of the non-linear velocity field and $f$ may not be directly proportional. \citet{chapman22} investigated the effect of varying $\gamma_f$ on the correlation function and identified a scale of $\sim 7\,h^{-1}\,{\rm Mpc}$ as the transition between the quasi-linear and non-linear regimes, so they restricted their measurement of $f\sigma_8$ to between $7-60\,h^{-1}\,{\rm Mpc}$ to isolate the linear signal when using a single scaling parameter.

We improve on the \citet{chapman22} emulator using the method described in Sec.\ref{sec:lin-vel} to model the linear and nonlinear velocity components. In order to apply this new method we require access to the initial conditions of the simulation, which are not publically available for the Aemulus suite of simulations \citep{derose_aemulus1} used by the \citet{chapman22} emulator. For our new emulator we use the AbacusCosmos suite of simulations \citep{abacus}, with available first-order initial conditions generated from the zeldovich-PLT\footnote{https://github.com/abacusorg/zeldovich-PLT} code. AbacusCosmos consists of 40 variable cosmology 1100 $h^{-1}\,{\rm Mpc}$ simulation boxes with $1440^3$ particles that we use to train the emulator, as well as 20 simulation boxes at the Planck 2015 cosmology \citep{planck2015} that are used for testing. Since the AbacusCosmos and Aemulus suites are similar in terms of number of boxes, box size, and number of particles we use the same method to estimate the emulator uncertainty as \citet{zhai_aemulus3}, adapted to the boxes available in AbacusCosmos. We use the 20 AbacusCosmos boxes with Planck cosmology to estimate the sample variance, and assess the performance of the emulator throughout the cosmological parameter space by retraining the emulator with one variable cosmology box excluded at a time, and comparing emulator predictions to measurements from the excluded box.

\subsection{Isolating the linear signal}
\label{sec:lin-vel}

 In order to ensure that our results are not biased by the assumption that all components of the velocity will be scaled in the same way by a change in $f$ we split the velocity of halos into two components: a linear and a non-linear component. We scale each component by an independent parameter: $\gamma_l$ for the linear component and $\gamma_n$ for the non-linear component. If these parameters are constrained such that $\gamma_l=\gamma_n$ then all velocities are scaled by the same amount and the model reduces to the single scaling parameter, $\gamma_f$ used in \citet{chapman22}. The split is performed on halo velocities rather than galaxy velocities because the velocity bias of galaxies is implemented by other independent parameters in the emulator. Galaxies are assigned the velocity of their host halo with an additional velocity term calculated as $\sigma_{\rm gal}=v_{\rm gal}\sigma_{\rm halo}$, where $v_{\rm gal}$ is the velocity bias parameter for that galaxy type ($v_{\rm bc}$ and $v_{\rm bs}$ for centrals and satellites respectively), and $\sigma_{\rm halo}$ is the velocity dispersion of the halo calculated from its mass using the virial theorem. The additional velocity term is calculated independently of the velocity scaling by $\gamma_l$ and $\gamma_n$ so that it is controlled entirely by $v_{\rm bc}$ and $v_{\rm bs}$. This choice reduces the degeneracy between the velocity scaling and velocity bias parameters while still allowing for sufficient freedom in the model to address both a change in the growth rate and the presence of velocity bias \citep{Guo:2015}.

The challenge of this new model is determining what component of the velocity is linear at late time. While this is difficult to do for the halo velocities, we can make use of the fact that the initial conditions of the emulator provide a method for calculating particle linear velocities, which can then be combined to provide an estimate of the linear velocity of the halo. The AbacusCosmos initial conditions were generated by calculating Zel'dovich approximation displacements for a grid of particles at $z=49$ using the zeldovich-PLT code. The Zel'dovich approximation provides a first order calculation of the displacements and velocities of particles, so $z=49$ is chosen as an arbitrarily large redshift where the motion of particles will very closely follow linear theory. We can use these initial particle linear velocities to predict the particle linear velocities at the $z=0.7$ simulation slice by evolving them using the linear theory prediction for the amplitude of the velocity field,
\begin{equation} \label{eq:lin-vel}
    \boldsymbol{v_k} = \frac{i\boldsymbol{k}}{k^2}Ha\delta_{\boldsymbol{k}}f(\Omega_m).
\end{equation}

The velocity scaling of the initial conditions is simply the ratio of Eq.~\ref{eq:lin-vel} between the redshift of the initial conditions and the desired final redshift,
\begin{equation} \label{eq:vel-scaling}
    \boldsymbol{v}(z_{2}) = \frac{Haf\sigma_8(z_{2})}{Haf\sigma_8(z_{1})} \boldsymbol{v}(z_{1}).
\end{equation}

We define the non-linear velocity as all components of the total velocity not included in the linear velocity, and calculate it by subtracting the linear velocity vector from the total velocity vector. By separately scaling the linear velocity by $\gamma_l$ and the non-linear velocity by $\gamma_n$ we allow for the non-linear velocity of the data to deviate from the $\Lambda {\rm CDM}$ expectation of the simulations without biasing the value of $f\sigma_8$ we infer from $\gamma_l$. $\gamma_l$ and $\gamma_n$ will, in general, be correlated with each other. For example, this will be true for quasi-linear velocity evolution that happens along the direction of the linear velocity. 

\subsection{Smoothing the linear velocity field}
\label{sec:smoothing}

\begin{figure*}
	\includegraphics[width=\textwidth]{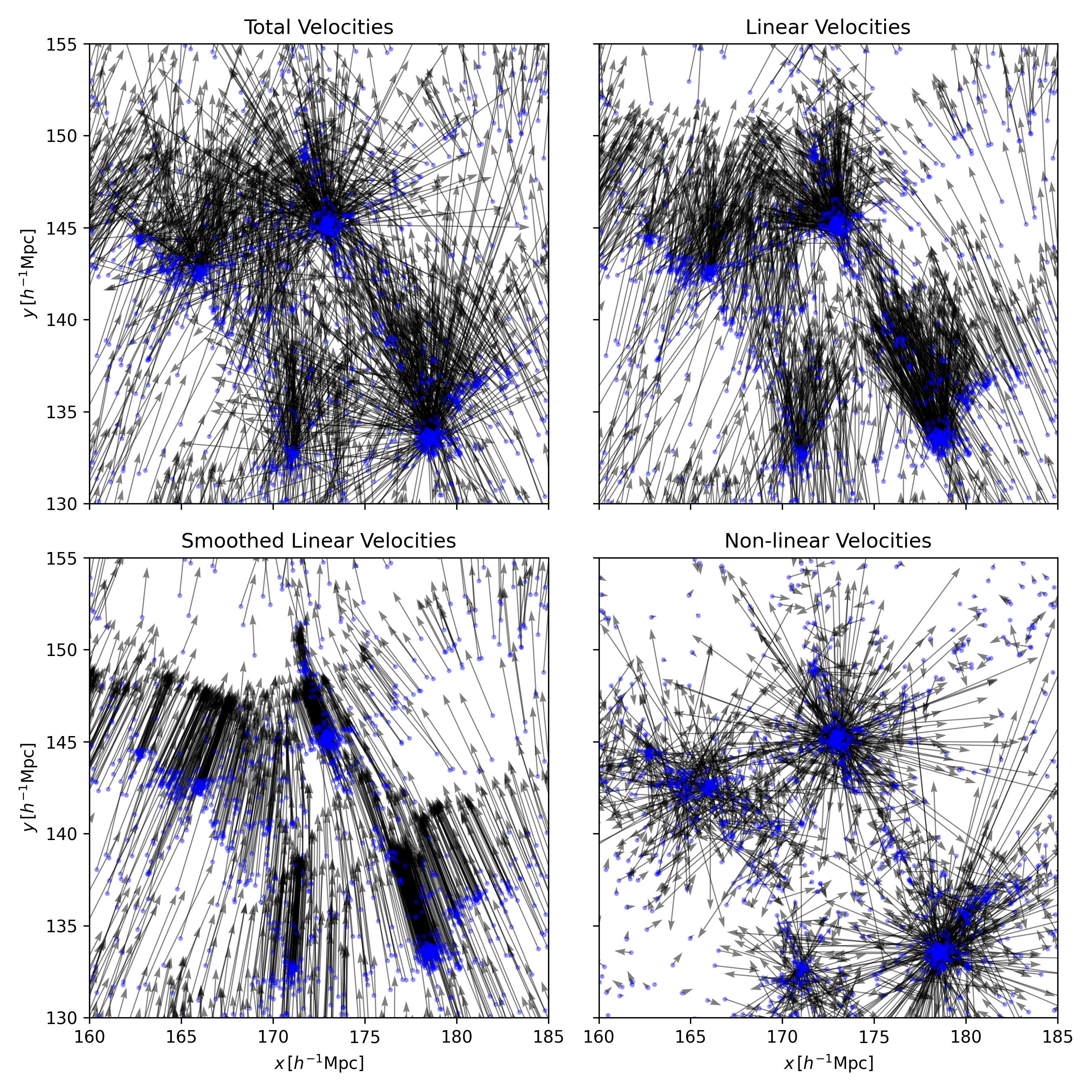}
    \caption{A slice of one of the Abacus Planck boxes showing the particle positions and velocities. Blue points show the position of particles from a uniform 10\% down sampling, and black arrows show the velocities of the particles where the size of the arrow is proportional to the amplitude of the velocity. \textit{Upper left:} The total particle velocity. \textit{Upper right:} The linear velocity calculated from the initial conditions. \textit{Lower left:} The smoothed linear velocity calculated using a tophat smoothing kernel with radius 5 $\mhmpc$. \textit{Lower right:} The non-linear velocity component, calculated as the difference between the total velocity and the linear velocity.}
    \label{fig:particle-vel}
\end{figure*}

Pairs of galaxies with small separation in collapsed objects have lost all dependence on the initial linear velocities. This approximately occurs at shell crossing and means that our split into linear and non-linear components is ineffective on such scales - a portion of the velocity ascribed to non-linear motion simply cancels out the linear one (see Appendix \ref{app:pairwise-vel}). In an extreme situation, if two objects are located sufficiently close to each other along the line of sight and have a large enough infall velocity, the shift in position in redshift space reverses the orientation of the pair along the line of sight. In this situation scaling the velocity will increase the pair separation, leading to damping of the correlation function. 
We therefore elect to smooth the particle linear velocity field around the shell crossing scale, which from our previous analysis we know to occur at approximately $5\mhmpc$. This smoothing reduces the pairwise linear velocity of nearby objects, transferring the component of the velocity that provokes shell crossing to what we have termed the 'non-linear' component, since total velocity is still conserved. Meanwhile, the linear pairwise velocity of more distant objects is unaffected, preserving the signal we wish to extract with our linear velocity scaling parameter.

To illustrate the smoothing effect we use a projected 5$\mhmpc$ thick slice of the Abacus Planck 00-0 box to demonstrate the arrangement of the different particle velocity components in a high density region, shown in Fig.~\ref{fig:particle-vel}. While the velocity of field particles is largely unchanged between total, linear, and smoothed linear velocities, the behaviour of particles in the cluster differs greatly. The unsmoothed linear velocity displays a distinct preferred direction when compared to the total velocity, however some scatter persists. The smoothed velocity is significantly more collimated so that close particles will maintain their separation in redshift space, as intended. The non-linear velocities show the difference between the total velocity and smoothed linear velocities. As expected, the non-linear velocities are significantly larger in collapsed structures compared to the field, and do not show an obvious preferred direction.

Our process of smoothing and assigning halo velocities is as follows. First, we construct a 3D grid with side length $1 \mhmpc$ over the simulation box, and assign to each grid cell a linear velocity equal to the mean linear velocity of the particles contained within the cell. Next, we smooth the grid using a 3D spherical tophat kernel of radius $5 \mhmpc$, equally weighting each grid cell. Finally, halos are assigned the smoothed linear velocity of the cell they inhabit. The smoothing radius of $5 \mhmpc$ was chosen to match the approximate scale found in \citet{chapman22} where increasing the velocity scaling parameter, $\gamma_f$, transitioned from amplifying the monopole to damping the monopole. A tophat kernel was chosen because of the small width of this transition, and because it reduces the number of calculations required for the smoothing compared to other possible kernel choices, such as a Gaussian kernel. The grid spacing was chosen to balance the resolution of the grid and the memory requirements of the computation. Testing these choices is discussed below.

\begin{figure*}
	\includegraphics[width=\textwidth]{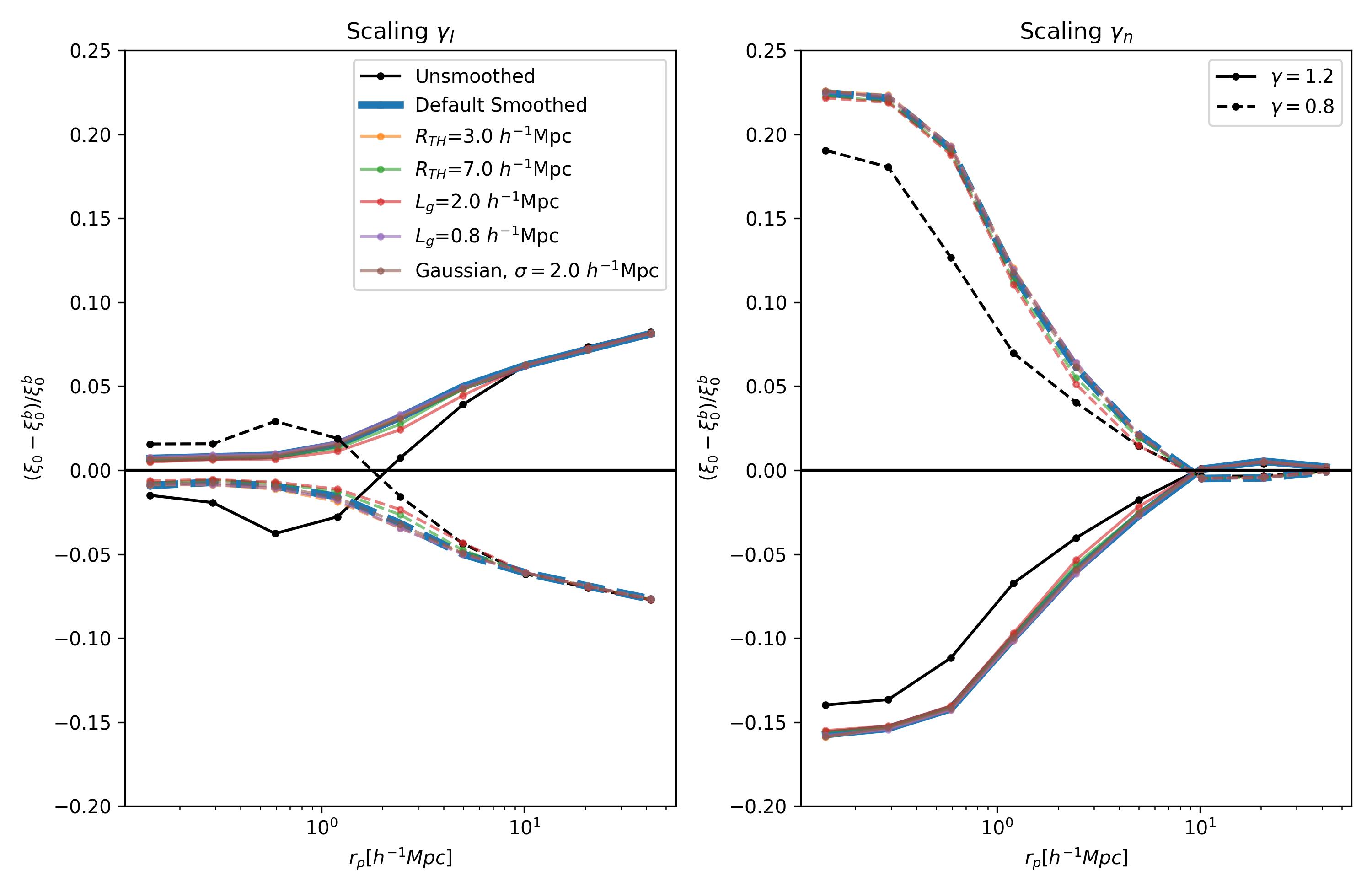}
    \caption{The mean change in the monopole of the redshift space halo correlation functions after velocity scaling from the 20 Planck cosmology boxes. The left panel shows the effect of scaling $\gamma_l$, while the right panel shows the effect of the scaling $\gamma_n$. Solid lines show the effect of scaling by $\gamma=1.2$, while dashed lines show the scaling by $\gamma=0.8$. The black lines show the result using the unsmoothed linear velocity, while the thick blue line shows the result of our fiducial smoothing method; a tophat kernel with radius 5 $\mhmpc$ on a grid of side length 1 $\mhmpc$. Faint coloured lines show the results of variations on the smoothing method. The orange and green lines show the results of varying the tophat smoothing radius to 3 and 7 $\mhmpc$ respectively while keeping the grid size fixed, while the red and purple lines show the result of varying the grad size to 2 and 0.8 $\mhmpc$ while keeping the smoothing kernel fixed to the fiducial method. Finally, the brown line shows the result of smoothing using a Gaussian kernel with standard deviation 2 $\mhmpc$ on a 1 $\mhmpc$ grid.}
    \label{fig:vel-smoothing-cf-check}
\end{figure*}

In Fig.~\ref{fig:vel-smoothing-cf-check} we investigate the effect of scaling the smoothed halo linear velocity on the monopole of the halo correlation function and compare to the results of scaling the unsmoothed halo velocities. For the unsmoothed linear velocity field we define the linear halo velocity as the mean linear velocity of the constituent particles. Scaling both the smoothed and unsmoothed velocities has a nearly identical effect on the large scales of the monopole for both the linear velocity scaling parameter, $\gamma_l$, and the non-linear velocity scaling parameter, $\gamma_n$. This result is expected since the velocity smoothing primarily affects the pairwise velocity of small separation objects by construction, and desired because the large scale behaviour follows the expectation from linear theory in that the amplitude of the monopole is proportional to $f$, and scaling up the velocities increases the amplitude of the correlation function. However, around $\sim 2\mhmpc$ scaling up the unsmoothed linear velocities changes behaviour and damps the monopole due to the shell crossing issue discussed above. Scaling up the smoothed linear velocity increases the amplitude at all scales, although the effect is reduced below the smoothing scale. This matches our desired behaviour for the linear velocity field, which was visualized in Fig.~\ref{fig:particle-vel}, that close pairs that have already collapsed maintain their separation as the linear growth rate is increased, rather than being spread apart. When scaling $\gamma_n$ the effect is similar for both methods of calculating the velocity components, although the smoothed velocity field shows a greater change in amplitude. The quadrupole is not included in this plot because the change in sign makes these trends more difficult to see intuitively, but the same behaviour of the scaling parameters is seen in quadrupole as displayed in the monopole. The projected correlation function is largely insensitive to the radial velocity by construction, and the difference between smoothed and unsmoothed velocities is insignificant.

Fig.~\ref{fig:vel-smoothing-cf-check} also shows the results of varying the parameters used to smooth the linear velocity field. Faint, coloured lines show the effects of smoothing using a tophat radius of $3.0 \mhmpc$ and $7.0\mhmpc$ instead of the default $5.0\mhmpc$, using a grid of side length $2.0 \mhmpc$ or $0.8 \mhmpc$ instead of the default $1.0\mhmpc$, and of using a Gaussian kernel with standard deviation $2.0\mhmpc$. In all cases the effect is quite similar to our default choice of parameters at all scales and for both scaling parameters, indicating that our smoothing method is robust to varying these choices.

\subsection{Testing the improved emulator}

We validate our emulator by performing an MCMC fit to a subsample of the measurements of the Planck 2015 boxes used for determining the emulator uncertainty. We randomly select 10 test HOD models and measure the redshift space galaxy correlation for all 20 simulation boxes with line-of-sight along each of the three axes, giving a total of 60 measurements. We average the results of these 60 measurements for each HOD model and fit the data using our improved emulator. For the covariance matrix we use our data covariance matrix, scaled along the diagonal to match the volume of the mock measurements without modifying the correlation structure. While the true effective volume of our measurement will be between 20-60 simulation boxes because we use 20 independent boxes each measured along three independent lines-of-sight, we choose a volume of 20 simulation boxes as our fiducial amount to be conservative.

For all 10 models we recover the known value of $\gamma_l$ and the expected value of $f\sigma_8$ to within the 68\% confidence interval. This is expected given our conservative choices for the emulator uncertainty, which lead to slightly inflated confidence intervals while ensuring that our parameter inference is not biased. Likewise the known cosmological and HOD parameters are recovered for the majority of the models. The HOD parameters that are least often recovered are $\log M_{\rm cut}$, $\sigma_{\log M}$, and $f_{\rm max}$, however none are degenerate with our key cosmological parameters and there is no significant impact on the $f\sigma_8$ constraints, so there is no concern for our measurement of the eBOSS data.

We also investigate the scale dependence of the constraints from the 10 test HOD models. For each model we perform a fit to the full separation range of the model, $0.1-60\mhmpc$, as well as four additional fits restricted to the separation ranges $0.1-7\mhmpc$, $0.8-7\mhmpc$, $0.8-60\mhmpc$, and $7-60\mhmpc$, matching the methodology used to test the data in Sec.~\ref{sec:scales-probes}. For each model we find all separation ranges give a mutually consistent value of $f\sigma_8$ at the $1\sigma$ level, with approximately half of the models showing a slight offset between the $0.1-7\mhmpc$ and $0.8-7\mhmpc$ results and the remaining separation ranges. The offset is equally likely to occur to larger and smaller values and is within the measurement uncertainty, so it is not a concern for our cosmological inference.

Finally, we validate our entire pipeline using a subhalo abundance matching (SHAM) mock constructed from the Uchuu\footnote{http://skiesanduniverses.org/Simulations/Uchuu/} simulation (\citealt{Ishiyama_2021}). Uchuu is a $(2000 \mhmpc)^3$, $12800^3$ particle simulation using the Planck2015 cosmology and a mass resolution of $m_p=3.27\times10^8 \,h^{-1}M_\odot$. Using a different galaxy halo connection model and simulation is a necessary test of the robustness of our model in order to be able to confidently apply it to the eBOSS data. Fitting the correlation function of the SHAM mock using our new emulator we are able to recover the known cosmological parameters within the 68\% confidence interval for all parameters, and find all well constrained HOD parameters to be within their respective prior ranges. We recover $\gamma_l=1.00\pm0.08$ and $\gamma_n=0.90\pm0.14$, both consistent with their expected values of 1 since the mock contained a $\Lambda {\rm CDM}$ growth rate and no velocity scaling.

%% file: 3-results.tex
\subsection{eBOSS LRG Sample}

We fit our new emulator model to the extended Baryon Oscillation Spectroscopic Survey (eBOSS) \citep{dawson_sdss-iv_2016} luminous red galaxy (LRG) sample analyzed in \citet{chapman22}. Targets were selected for the eBOSS LRG sample \citep{Prakash:2016} from a combination of SDSS DR13 photometry \citep{Albareti:2017} and infrared observations from the WISE satellite \citep{Lang:2016}. Spectroscopic observations were made using the BOSS spectrographs \citep{Smee:2013} mounted on the 2.5-meter Sloan telescope \citep{Gunn:2006}. The eBOSS LRG sample consists of 174 816 objects over 4242 ${\rm deg}^2$ in the redshift range $0.6<z<1.0$. The sample has an effective volume of 1.28 ${\rm Gpc}^3$, an effective redshift of $z=0.737$, and a peak number density of $n=1\times 10^{-4}\, ({\rm Mpc}^{-1}h)^3$ \citep{Ross20a}.

We apply the standard eBOSS weights to the data, which correct for variations in obtaining reliable redshifts and observational contaminants, as well as optimizing the signal obtained from the data. We also apply the pairwise inverse probability weights combined with angular upweighting (PIP+ANG) \citep{Bianchi:2017, Percival:2017} calculated in \citet{Mohammad:2020} to correct the fibre collision issue. Fibre collision occurs when the physical size of the fibres prevents simultaneously targeting multiple close objects within a single pointing of the instrument, leading to a biased sample that is particularly concerning for small scale observations. \citet{Reid:2014} identified fibre collision as the the most significant issue for analyzing small scale clustering of SDSS data. Most analyses use an approximate correction that involves transferring the weight from the missing object to a nearby observed object. This type of correction approximately recovers the true clustering on large scales, however the performance degrades on smaller scales and all information below the fibre collision scale is lost. By contrast PIP weights are theoretically unbiased on all scales, allowing for a full recovery of the true clustering. \citet{Mohammad:2020} calculated PIP+ANG weights for all three eBOSS samples, which we apply when measuring the eBOSS LRG clustering.

We measure the clustering using the two-point correlation function, which represents the excess probability of finding two galaxies at a given separation compared to if the sample was randomly distributed. We calculate the correlation function using the Landy-Szalay estimator, which has been shown to be the least-bias and least-variance estimator \citep{landy93}. We use a random catalogue matching the angular and radially distribution of the LRG sample with a factor of 50 times more points in order to reduce the impact of shot noise. The difference in the number of data and random points is taken into account in the normalization of the pair counts.

To reduce the number of bins, we compress the full 2D correlation function in two ways. We use the first two even multipoles of the correlation function, which contain most of the RSD information. It is common to also use the next even multipole, the hexadecapole, in RSD analyses. However, due to the increased noise in the hexadecapole and required increase in complexity of the emulator we choose to exclude it. The second compression method is the projected correlation function,

\begin{equation}
    w_p\left(r_\perp\right) = 2\int_{0}^{r_{\parallel, {\mathrm{max}}}}\xipp dr_{\parallel}, \label{eq:wp}
\end{equation}

where $r_{\perp}$ and $r_{\parallel}$ are the normal and parallel to the line-of-sight components of the pair separation, $\boldsymbol{s}$. We limit the integral to a $r_{\parallel, {\mathrm{max}}}=80\,h^{-1}{\rm Mpc}$, which is sufficient to remove the majority of the RSD signal. While not sensitive to RSD, $w_p$ is useful for constraining the HOD parameters of our model because it has different parameter degeneracies than the multipoles.

We bin both $r_{\perp}$ and $s$ in 9 logarithmically spaced bins between $0.1-60\mhmpc$, while $r_{\parallel}$ and $\mu$ are binned using linear bins of width $\Delta r_{\parallel}=1\mhmpc$ and $\Delta\mu=0.1$. These measurement statistics and binning schemes are matched to those used in our emulator. The spacing of the separation bins is chosen to sample a range of scales commonly excluded in traditional measurements, while limiting the number of bins in order to reduce the training complexity of the Gaussian process emulator. Additionally, in order to ensure a match between the model and the data we scale the separations of the model correlation functions by the Alcock-Paczynski parameters (\citealt{AP79}) to account for the difference between the cosmology of the fit and the fiducial cosmology used to convert redshifts to distances for the data (see Sec. 3.6 of \citealt{chapman22}).

We estimate the covariance of our measurements using jackknife resampling of the data in 200 angular regions. To select the regions we apply equal area angular tiles to the data footprint, and remove the lowest occupation regions to arrive at 200 equally sized and equally weighted angular regions. The covariance is then estimated by removing regions one at a time, measuring the clustering of the remaining regions, and comparing to the measurement over the whole survey according to

\begin{equation}  \label{eq:jackknife}
    C_{i,j} = \frac{n-1}{n}\sum_{k}^{n}(\xi_{i,k} - \bar{\xi}_i)(\xi_{j,k} - \bar{\xi}_j),
\end{equation}

where the $i,j$ indices are over the elements of the data vector, n=200 is the number of jackknife regions, and $k$ is an index over the jackknife realisations. We then rescale the diagonals of the covariance matrix to account for the difference in volume between the 200 jackknife regions and the full catalogue while preserving the correlation structure. For more details, see \citet{chapman22}. In this work we also apply the $v_{\rm match}$ weighting scheme detailed in \citet{mohammad22}, which corrects the ratio of auto-pairs and cross-pairs removed by the jackknife sampling for a low density of galaxies. Applying these weights we find a minor reduction in the covariance of large separation bins, matching what was seen in \citet{mohammad22}, although there is no significant change to the correlation between different separation bins.

\subsection{Headline results}

\begin{figure*}
	\includegraphics[width=\textwidth]{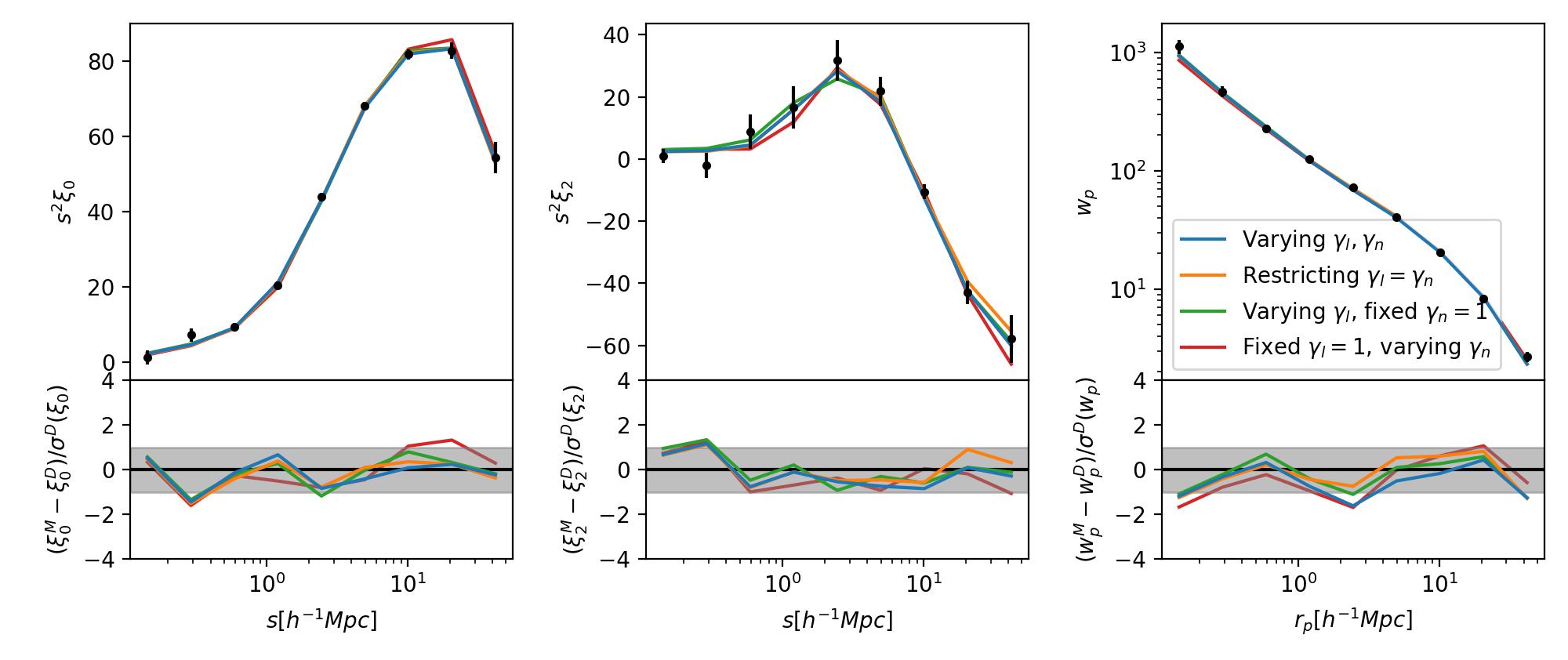}
    \caption{Best fit models compared to the eBOSS LRG measurement data for several choices of scaling parameters. Our baseline fit, allowing both $\gamma_l$ and $\gamma_n$ to vary, is shown in blue. A single velocity scaling parameter model, constrained so that $\gamma_l=\gamma_n$ and equivalent to the model used in \citet{chapman22}, is shown in orange. The green line shows the result of allowing $\gamma_l$ to vary while fixing $\gamma_n=1$, and the red line shows the result of allowing $\gamma_n$ to vary while fixing $\gamma_l=1$. The left, centre, and right columns show the monopole, quadrupole, and projected correlation function respectively. The top row of panels directly compares the model to the data, while the lower row shows the difference between model and data in units of the data uncertainty, with the grey-shaded region indicating the $1\sigma$ region.}
    \label{fig:best-fit}
\end{figure*}

\begin{figure*}
	\includegraphics[width=\textwidth]{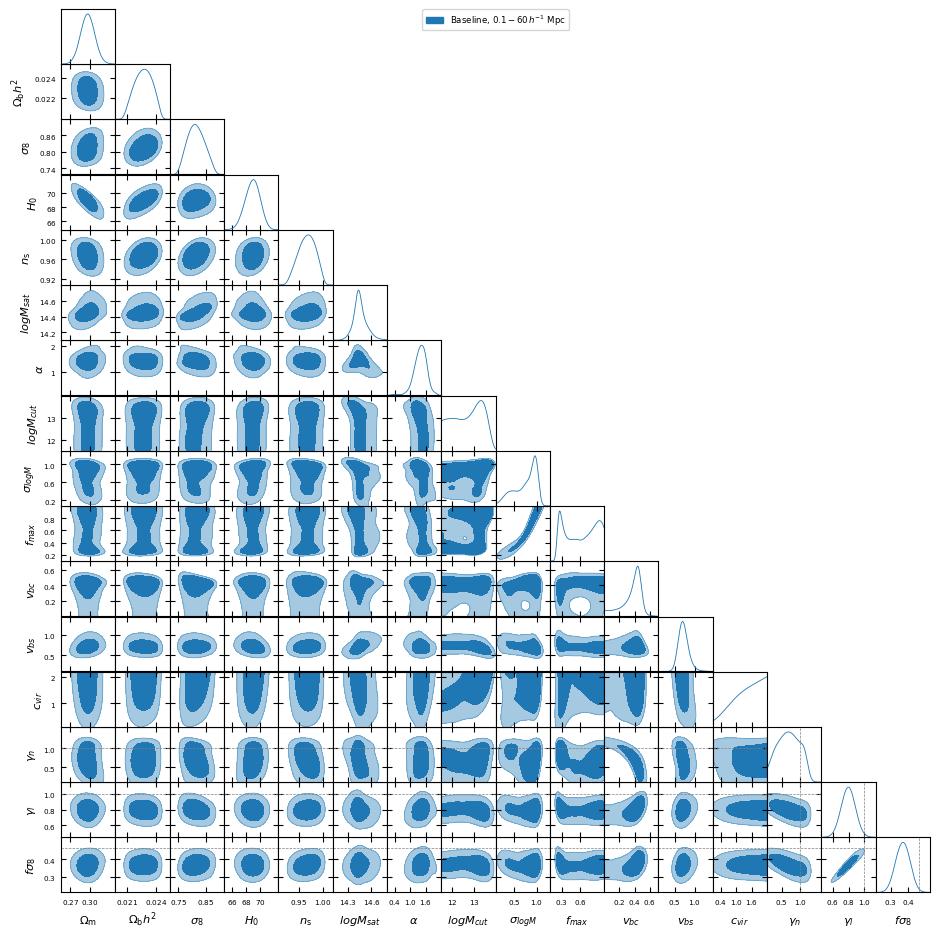}
    \caption{1D and 2D contours of the parameters used in our baseline fit of the eBOSS LRG $\xi_0$, $\xi_2$, and $w_p$ in the separation range $0.1-60 \mhmpc$. The constraint on $f\sigma_8$ is calculated as $f\sigma_8=\gamma_l f_{\Lambda {\rm CDM}}\sigma_8$. The dashed lines highlight $\gamma_l=1$ and $\gamma_n=1$, which would indicate that no velocity scaling is needed to match the data to the $\Lambda CDM$ expectation of the emulator.}
    \label{fig:triangle-plot}
\end{figure*}

Our analysis of the eBOSS sample with the new velocity-split emulator yields a value of $f\sigma_8(z=0.737)=0.368\pm0.041$, with $\chi^2=16.2$ from 27 data points and 15 free parameters. This value is 2.3-$\sigma$ below the expectation for a $\Lambda{\rm CDM}$ universe with the Planck 2018 cosmology, and an increase in tension from the 1.4-$\sigma$ offset found in \citet{chapman22}. \citet{chapman22} found $f\sigma_8(z=0.737)=0.408\pm0.038$ when using a single velocity scaling parameter and limiting their measurement scales to $7-60\mhmpc$, so this increase in tension is caused by a shift to a lower value of $f\sigma_8$ rather than an increase in precision, although the two results are mutually consistent.

In Fig.~\ref{fig:best-fit} we compare the best fit models for various choices of scaling parameters to the eBOSS data. All models are able to accurately fit the data on all scales, although our baseline model of allowing both $\gamma_l$ and $\gamma_n$ to vary results in the lowest $\chi^2$ value. The performance is 
slightly improved over a single scaling parameter model in the large scales of the quadrupole and projected correlation function. This is likely caused by improved flexibility in simultaneously fitting the smallest and largest measurement scales by decoupling the scaling of the velocity terms, with the non-linear velocity dominating on the smallest scales and the linear velocity dominating on the largest scales. It should be noted that while the fixed $\gamma_l=1$ model is restricted to match the linear velocity amplitude expected for a $\Lambda {\rm CDM}$ universe from the AbacusCosmos simulations, it does not indicate agreement with the value of $f\sigma_8$ expected from the Planck 2018 observations because $\Omega_m$ and $\sigma_8$ are still allowed to vary. That model results in values of $\sigma_8=0.791\pm0.027$ and $f\sigma_8=0.450\pm0.016$, with $\chi^2=20.4$.

Our fit to the data only weakly constrains the amplitude of the non-linear velocity field, giving a value of $\gamma_n=0.694\pm0.29$, where a value of $\gamma_n=1$ indicates agreement between the data and the expectation for a $\Lambda {\rm CDM}$ universe from the model. This constraint is limited by the lower edge of the prior at  $\gamma_n=0.5$, but does show a clear preference for $\gamma_n<1$. The poor constraint is likely due to the lower magnitude of the non-linear velocity field compared to the linear velocity field (see Appendix \ref{app:pairwise-vel}), as well as the degeneracy between $\gamma_n$ and $v_{\rm bc}$ (Fig.\ref{fig:triangle-plot}). It may also indicate that our parameterization of $\gamma_n$ needs further refinement in order to fully describe the behaviour of the actual non-linear velocity field. We have implemented $\gamma_n$ as a uniform scaling for all components of the velocity that do not match the initial linear velocity. It is possible that there are multiple contributions to the non-linear velocity, requiring a more nuanced parameterization to capture the deviations between the data and the best fit $\Lambda {\rm CDM}$+HOD model. Non-linear velocity scaling is also not necessarily uniform for all galaxies, and may be dependent on characteristics such as galaxy mass and environment. Investigating these alternatives is a possible avenue for future research. While this result is independent of our cosmological constraint by construction, it does indicate that non-linear velocities in the data are lower than those generated by combining our HOD model with a CDM-only simulation.

\subsection{Testing the dependence on the data fitted}
\label{sec:scales-probes}

\begin{figure*}
    \centering
    \begin{subfigure}[b]{\columnwidth}
    	\includegraphics[width=\columnwidth]{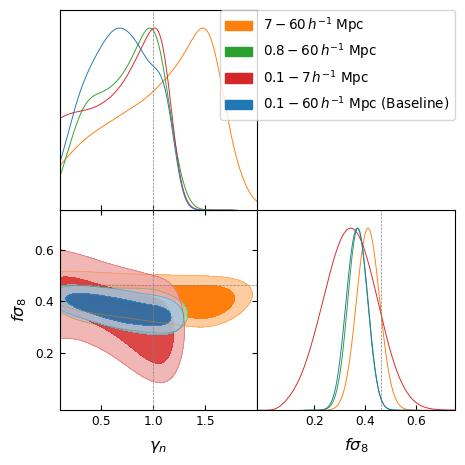}
        \label{fig:scales}
    \end{subfigure}
    \hfill
    \begin{subfigure}[b]{\columnwidth}
    	\includegraphics[width=\columnwidth]{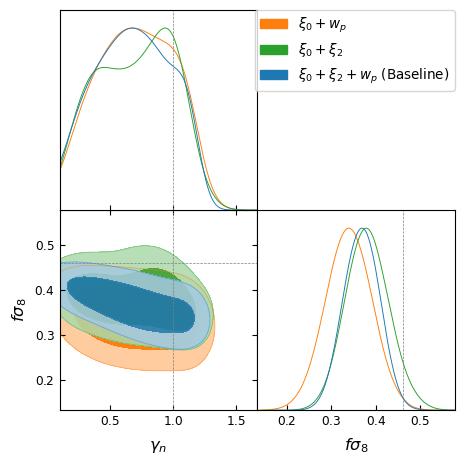}
        \label{fig:probes}
    \end{subfigure}
        \caption{2D and 1D marginalized constraints on $\gamma_n$ and $f\sigma_8$ for fits to different scales and measurements. \textit{Left:} Constraints from the three largest separation bins (orange), six largest separation bins (green), six smallest separation bins (red), and all nine separation bins (blue) for all three measurements. The dotted line shows the value of $f\sigma_8$ expected from the Planck 2018 results assuming a $\Lambda$CDM cosmological model. \textit{Right:} Constraints from the joint fit to the monopole and projected correlation function (orange), monopole and quadrupole (green), and all three measurements (blue).}
        \label{fig:scales-probes}
\end{figure*}

A key motivating factor for constructing our new velocity-split model was the scale dependence observed when using a single velocity scaling parameter by \citet{chapman22}. That analysis found that fitting to various measurement scales found lower values of $f\sigma_8$ at smaller scales, although all measurements were consistent with each other and below the expectation for a $\Lambda{\rm CDM}$ universe with Planck 2018 cosmology. Using our updated emulator we find the smallest measurement scales to be in better agreement with the larger scales of our analysis. A small offset still exists between the quasi-linear scales and transition scales, as shown in Fig.~\ref{fig:scales-probes}. A comparison of the constraints on $f\sigma_8$ using various measurement scales between the new emulator and the result of the the single velocity scaling parameter emulator used in \citet{chapman22} is shown in Table~\ref{table:scales}.

\begin{table}
\centering
\begin{tabular}{|l|c|c|}
    \hline
    Measurement Scales & \citet{chapman22} & This Work \\
    \hline
    $0.1-7\,h^{-1}{\rm Mpc}$ & $0.334\pm0.061$ & $0.335\pm0.105$ \\
    $0.8-60\,h^{-1}{\rm Mpc}$ & $0.373\pm0.031$ & $0.368\pm0.041$ \\
    $7-60\,h^{-1}{\rm Mpc}$ & $0.408\pm0.038$ & $0.412\pm0.048$ \\
    $0.1-60\,h^{-1}{\rm Mpc}$ & $0.365\pm0.025$ & $0.368\pm0.041$ \\
    \hline
\end{tabular}
\caption{Comparison of $f\sigma_8$ constraints from different scales between the velocity-split emulator and a single velocity scaling parameter emulator.}
\label{table:scales}
\end{table}

This result follows our expectation for splitting the velocity parameters into $\gamma_n$ and $\gamma_l$. $\gamma_n$ is more important on the smallest scales, which are fit to the lowest velocity amplitude. Introducing an additional degree of freedom for the non-linear velocities through $\gamma_n$ reduces the tension in $\gamma_l$, but leaves the constraints from the scales around the transition from non-linear to quasi-linear ($\sim0.8-7\mhmpc$)  and quasi-linear scales largely unaffected. Our lower overall value of $f\sigma_8$ from the new analysis is caused by the inclusion of these transition scales, which also preferred a low value of $f\sigma_8$ in \citet{chapman22} but could not be definitively attributed to the linear signal until the introduction of our new model.

While the introduction of $\gamma_n$ and $\gamma_l$ has not fully removed the scale dependence of our measurement, it has significantly improved the agreement of different measurement probes, as shown in the right panel of Fig.~\ref{fig:scales-probes}. The results of fitting to the multipoles alone, the monopole with projected correlation function, and all three measurements are in close agreement. This result is a significant improvement over \citet{chapman22}, which found some tension between the different measurements due to the degeneracy between the combined velocity scaling parameter and $v_{bc}$.

\subsection{Comparison to previous emulator}

\begin{figure}
	\includegraphics[width=\columnwidth]{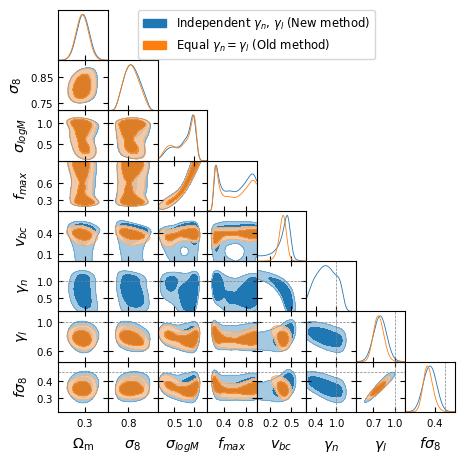}
    \caption{2D and 1D marginalized constraints of several key parameters from the eBOSS LRG data using the independent velocity-split scaling parameters introduced in this paper (blue) compared to the results from a single scaling parameter (orange), such as that used in \citet{chapman22}. Both fits were made using the updated emulator described in Sec.~\ref{sec:isolating}. For the single scaling parameter fit we constrain $\gamma_n=\gamma_l$ in order to mimic the effect of a single scaling parameter for all components of the velocity.}
    \label{fig:emulator-comp}
\end{figure}

In Fig.~\ref{fig:emulator-comp} we compare the constraints of key parameters between our new velocity-split model and the single velocity scaling parameter model used in \citet{chapman22}. To ensure an accurate comparison between the two methods we produce a new fit using our current emulator by setting $\gamma_l=\gamma_n$, which is equivalent to scaling all velocities by a single value. We find that all parameters are consistent between the two methods, with the most significant differences occurring in the velocity parameters $\gamma_n$, $\gamma_l$ and $v_{\rm bc}$, as expected. The new method slightly increases the uncertainty on $\gamma_l$, which follows through to the $f\sigma_8$ constraint, since splitting the velocity scaling parameters causes $\gamma_l$ to have a smaller impact on the fit, particularly at small scales. $\gamma_n$ and $v_{\rm bc}$ show significant degeneracy since they both contribute dispersive components to the galaxy velocity, and there are lesser degeneracies between $\gamma_l$ and $\gamma_n$, and $\gamma_l$ and $v_{\rm bc}$.

%% file: 4-discussion.tex
\begin{figure*}
	\includegraphics[width=\textwidth]{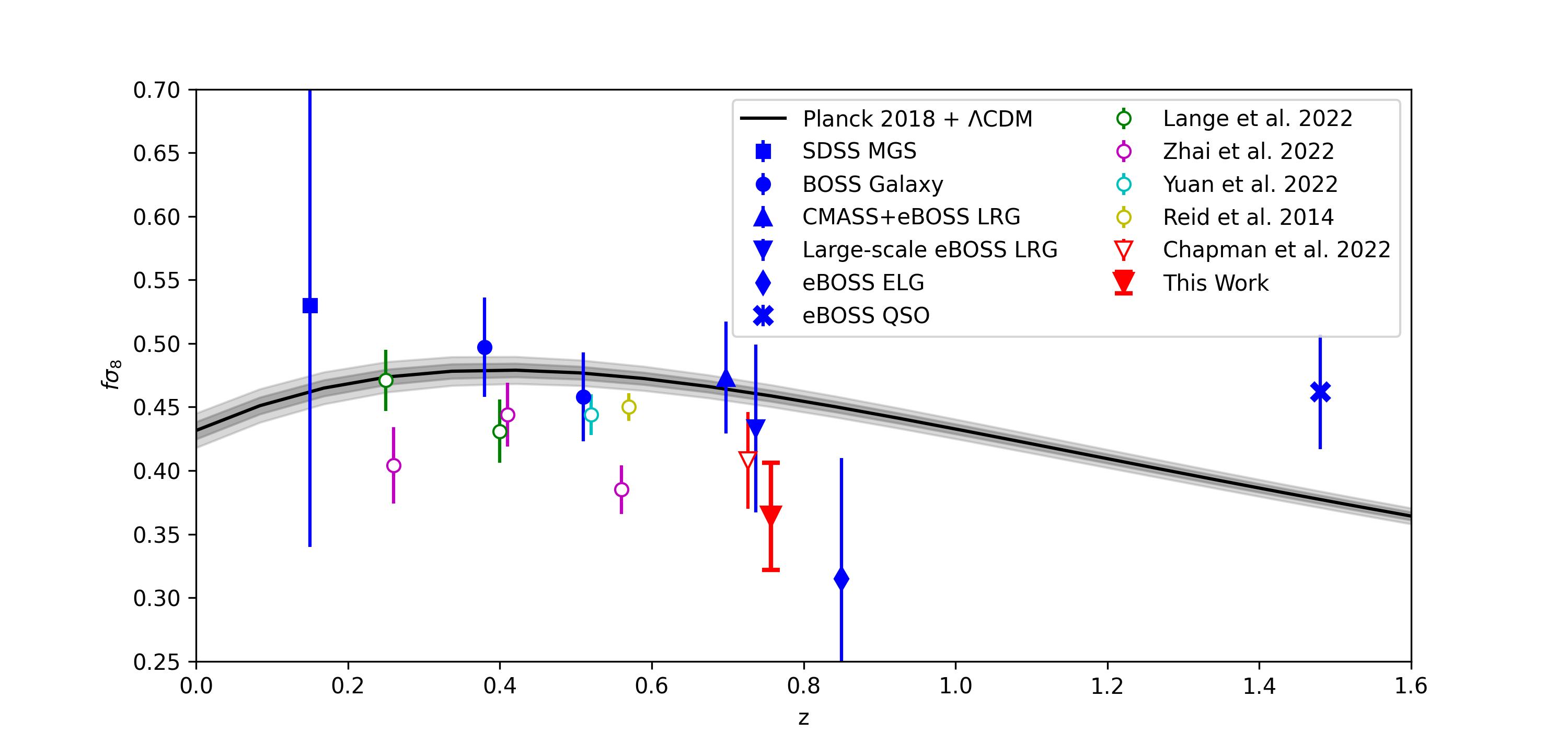}
    \caption{$f\sigma_8$ constraints from RSD measurements of SDSS samples. The blue points show the results of traditional large-scale analyses from the SDSS MGS \citep{Howlett_clustering_2015}, BOSS galaxies \citep{Alam:2017}, CMASS+eBOSS LRGs, eBOSS LRGs \citep{LRG_corr}, eBOSS ELGs \citep{demattia20a}, and eBOSS quasars \citep{neveux20a}. The results of this work are shown as the red solid triangle, while the red empty triangle shows the results from \citet{chapman22} using only the separation range $7-60\mhmpc$. Other coloured points show the results from various small-scale analyses of the BOSS galaxy samples that do not distinguish between linear and non-linear information (\citealt{Lange:2021} in green, \citealt{Zhai_2021} in magenta, \citealt{yuan22} in cyan, and \citealt{Reid:2014} in yellow). The black line shows the expected value of $f\sigma_8$ for a flat $\Lambda$CDM universe with best fit Planck2018 cosmology, with the shaded regions showing the 1 and 2$\sigma$ confidence regions. Measurements from the same galaxy sample are shifted slightly in the x-axis to avoid overlap.}
    \label{fig:sdss-results}
\end{figure*}

Using a emulator-based model with individual scaling parameters for the linear velocity, $\gamma_l$, and non-linear velocity, $\gamma_n$, we measure $f\sigma_8(z=0.737)=0.368\pm0.041$ from clustering between $0.1-60\,h^{-1}{\rm Mpc}$. \citet{chapman22} measured the same sample using an emulator with a single parameter scaling for the total velocity, but restricted their range of measurement to $0.7-60\,h^{-1}{\rm Mpc}$ in order to isolate the linear signal, and found $f\sigma_8(z=0.737)=0.408\pm0.038$. The shift to lower values in the updated emulator is caused by the inclusion of smaller scale clustering, and is very similar to the measurement from the same scales using the older emulator.

The consistency of the $f\sigma_8$ constraints between the two models gives confidence that our cosmological constraint is robust to the form of the velocity scaling. The advantage of the new model is that by isolating the linear signal we can now confidently extend our fitted data to small scales, which gives an increased tension with the expectation from Planck+$\Lambda {\rm CDM}$. By splitting the velocity scaling parameter to isolate the linear signal we can identify where the information for our constraint comes from, and be sure that we are optimally extracting linear information from the small-scale RSD signal without contamination from non-linear structure growth. This theory is borne out by the consistency between the results of the two emulators given the difference in modelling choices, which indicates that the non-linear velocities are not significantly affecting the linear measurements. Therefore, the most significant advancement of the new emulator is removing non-linear contamination as a potential source of systematic. In addition to the change in parameters, the older emulator was trained on the Aemulus simulation suite while the updated emulator was trained on the AbacusCosmos suite. The consistency between two different simulation suites, generated using different codes, indicates the reliability of the training data. Combined, these factors place severe limits on potential systematic biases in the analysis that could produce the low value of $f\sigma_8$ found from the data.

The results of both emulators, as well as other measurements of $f\sigma_8$ from SDSS galaxy samples are shown in Fig.~\ref{fig:sdss-results}. Our result is still consistent with the large scale analysis of the same sample at the $1\sigma$ level, but is now in $2.3\sigma$ tension with the expectation for a $\Lambda {\rm CDM}$ universe with a Planck2018 cosmology. There remains a consistent trend in small scale RSD measurements to lower values of $f\sigma_8$. This trend is now remarkable when considering the differences in modelling, data, and simulations between these analyses, shown in Table~\ref{table:small-scale-rsd}.

\begin{table*}
\centering
\begin{tabular}{|l|c|c|c|}
    \hline
    Analysis & Data & Simulations & Model \\
    \hline
    This Work & eBOSS LRG & AbacusCosmos & Emulator+$\gamma_l,\gamma_n$ \\
    \citet{chapman22} & eBOSS LRG & Aemulus & Emulator+$\gamma_f$ \\
    \citet{Lange:2021} & BOSS LOWZ & Aemulus & Cosmological Evidence Modelling \\
    \citet{Zhai_2021} & BOSS LOWZ+CMASS & Aemulus & Emulator+$\gamma_f$ \\
    \citet{yuan22} & BOSS CMASS & AbacusSummit & Constrained HOD Emulator \\
    \hline
\end{tabular}
\caption{Data, simulations, and models used by a variety of small scale RSD analyses.}
\label{table:small-scale-rsd}
\end{table*}

The consistency of these various small-scale RSD analyses leaves limited options to explain the tension with the $\Lambda {\rm CDM}$ expectation without modifying the cosmological model given the variations in data, simulations, and model. However, there are several common tools shared by all these analyses. All have models based on CDM-only simulations, with  an HOD model to connect galaxies to halos, and all are used to analyse data composed mainly of LRGs observed using the BOSS spectrograph. A non-cosmological solution could take the form of an overlooked systematic related to one of these three shared tools that is able to affect all analyses. However, it should be stated that each analysis has attempted to test these factors, and none have given evidence of an unknown systematic. In order to address these possible biases it is important to test the small scale clustering against simulations including baryonic physics \citep{Amon:2022}. Extending these analyses to DESI would also reduce the possibility of an observational bias since DESI uses a different target selection and significantly improved instrument \citep{desi1, desi2}. Measuring the DESI emission line galaxy (ELG) sample would be particularly interesting, since ELGs are expected to have a different HOD from LRGs, allowing an independent test of the HOD model. These factors could also explain the low value of $\gamma_n$ obtained in this analysis, which indicates a discrepancy between the model and the data in both the linear and non-linear velocity fields.

While our model has shown promise in isolating the linear signal, there remain a number of possible areas of improvement. The smoothing of the linear velocity field, while empirically motivated and tested, could be directly connected to a physical phenomenon \citep{Hollinger:2021}. The optimal smoothing scale is likely related to some physical characteristic of the density field, such as a the radius for shell crossing. The linear velocity parameter is poorly constrained, and several significant degeneracies exist between parameters in the fit. Refining these parameters could lead to more informative and precise results. Finally, while the model successfully separates linear growth and random motions, quasi-linear growth along the direction of the linear velocity remains a point of degeneracy between $\gamma_l$ and $\gamma_n$, and a potential bias in the model.

%% file: appendix.tex
\begin{figure*}
    \includegraphics[width=\textwidth]{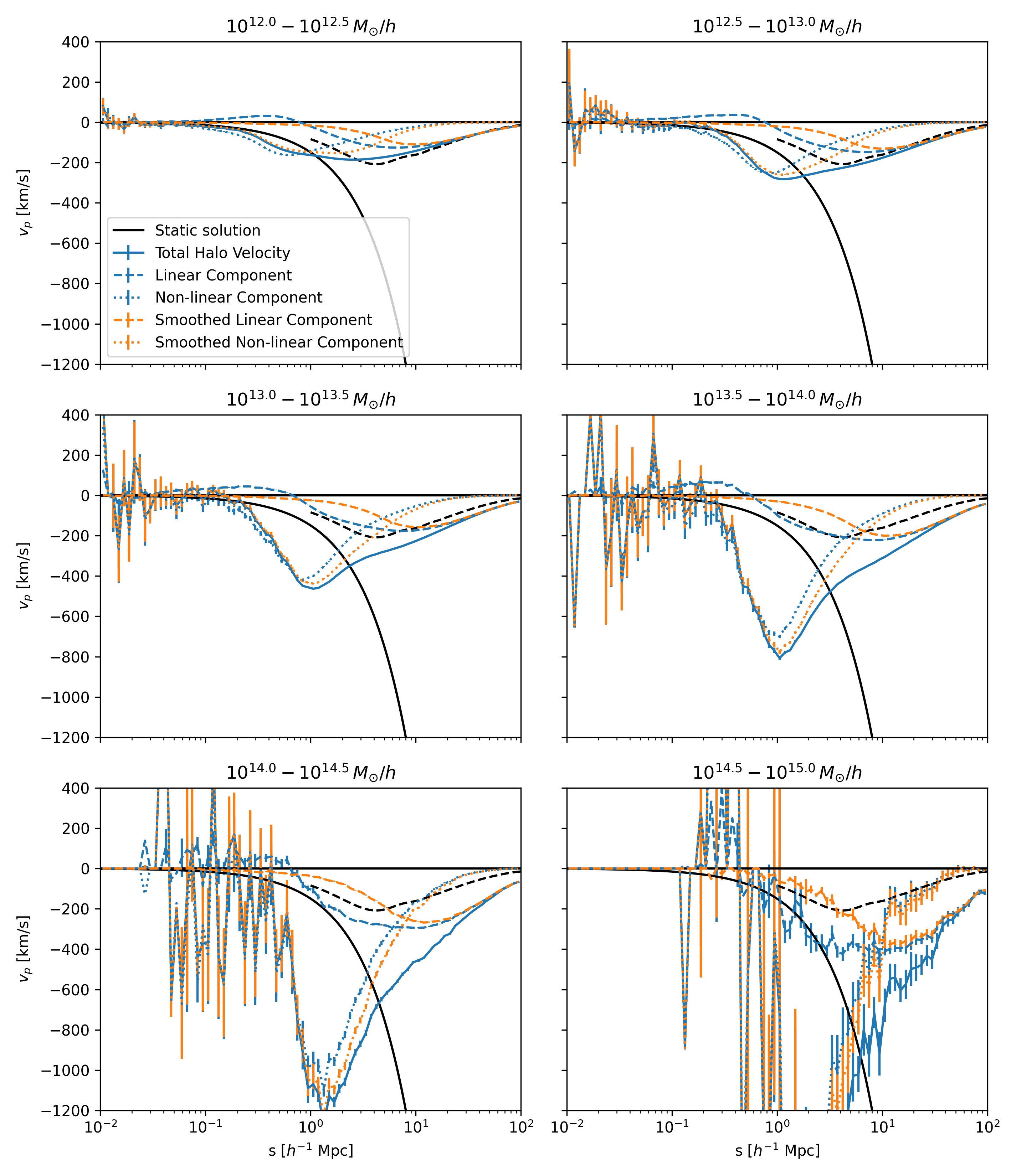}
    \caption{Mean pairwise velocities as a function of separation for the halos of the 20 Abacus Planck simulation boxes with box size 1100 $\mhmpc$, split into halo mass bins of width 0.5 dex from $10^{12}-10^{15} M_{\odot}/h$. The solid coloured line shows the total pairwise velocity, while dashed and dotted lines show the linear and non-linear components respectively. Blue lines show the linear and non-linear velocities calculated using the unsmoothed linear velocity, while orange lines show the smoothed linear and non-linear components. The solid black line shows the static solution and the dashed black line shows the theoretical linear theory prediction for the simulation cosmology.}
    \label{fig:pairwise-vel}
\end{figure*}

\begin{figure}
	\includegraphics[width=\columnwidth]{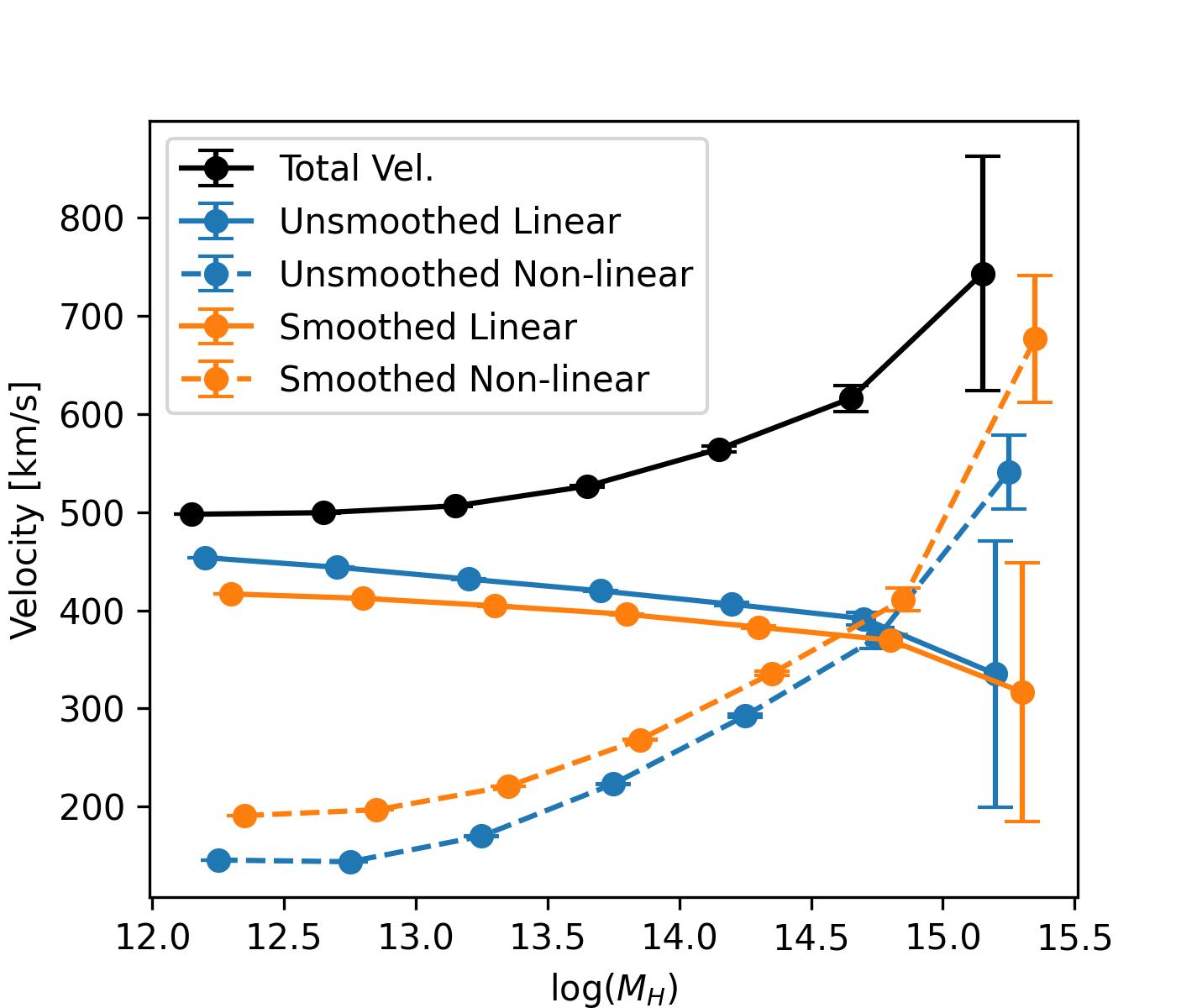}
    \caption{Mean halo velocity magnitude in mass bins of width 0.5 dex. The black line shows the mean total halo velocity, while solid coloured lines show the mean linear velocity magnitude and dashed coloured lines show the mean non-linear velocity magnitude. Blue lines represent the result using the unsmoothed linear velocity, while orange lines show the result after our fiducial velocity smoothing. The x position of points are offset slightly to prevent overlap.}
    \label{fig:vel-dist}
\end{figure}

In order to test our calculation of the linear velocity and to observe the relative impacts of the linear and non-linear components as a function of separation we examine the mean pairwise velocity of halos as a function of pair separation. We use the halos of the 20 boxes of the Abacus simulation suite with Planck 2015 cosmologies described in Sec.~\ref{sec:emulator}, at $z=0.700$. For each simulation we divide the halos into mass bins of width 0.5 dex from $10^{12}-10^{15} M_{\odot}/h$ and calculate the mean pairwise velocity for each mass interval, in 80 separation bins of equal logarithmic width from $10^{-2}-10^{2}\mhmpc$. We perform this calculation for both the unsmoothed and smoothed halo velocities (see Sec.~\ref{sec:smoothing}). These pairwise velocities are shown in Fig.~\ref{fig:pairwise-vel}. Also included in the plot is the static solution, which is the pairwise velocity required to maintain a constant proper separation in an expanding background, and the linear theory prediction for the pairwise velocity \citep{Fisher:1995, reid2011, belloso2012}. It should be noted that the linear theory prediction contains a bias factor that we have set to 1 for the calculation for all halo masses, leading to a difference in amplitude for the high mass halo bins.

At large separation the linear velocity calculated from the initial conditions is in good agreement with the total halo velocity as expected, providing a good test of the linear velocities calculated at high redshift as well as the scaling to the low redshift. The shape of the pairwise velocity is also a good match to the expectation from linear theory above $\sim20\mhmpc$. Below that scale the linear velocity begins to deviate form the total velocity and go towards 0 at small scales. Because the non-linear velocity is defined as the difference between the total velocity and the linear component this leads to a larger non-linear component, which peaks around $1\mhmpc$ after crossing the static solution before trending towards 0. The unsmoothed linear velocity becomes positive, indicating a pair increasing in separation, below $\sim0.5\mhmpc$. This behaviour is counter intuitive for structure growth and accounts for the unexpected effect of the scaling parameter on the correlation function at small scales shown in Fig.~\ref{fig:vel-smoothing-cf-check}. Smoothing the linear velocity field causes it to turn towards 0 below the smoothing scaling, without crossing to positive velocities at small separations.

We also investigate the magnitudes of the total, linear, and non-linear halo velocities as a function of halo mass in Fig.~\ref{fig:vel-dist} for the unsmoothed and smoothed velocity fields. Total velocity increases with halo mass, caused by an increase in non-linear velocity. The linear velocity dominates for small halo masses, but decreases slightly with halo mass. These trends are seen for both the unsmoothed and smoothed velocities, with the only significant difference being a slight decrease in the magnitude of the linear component from smoothing.